\documentclass[12pt]{article}
\usepackage[cp1251]{inputenc}
\usepackage[russian]{babel}
\usepackage[T2A]{fontenc}
\begin{document}
\begin{center} {\bf Geometrization of some quantum mechanics formalism} \end{center}
\begin{center}  О. А. Ol'khov* \end{center}
\begin{center} {\it N.N.Semenov Institute of Chemical Physics, Russian Academy of Sciences, Moscow} \end{center}

There were many attempts to geometrize electromagnetic field and
find out new interpretation for quantum mechanics formalism. The
distinctive feature of this work is that it combines geometrization
of electromagnetic field and geometrization of material field within
the unique topological idea. According to the suggested topological
interpretation, the Dirac equations for a free particle and for a
hydrogen atom prove to be the group--theoretical relations that
account for the symmetry properties of localized microscopic
deviations of the space--time geometry from the pseudoeuclidean one
(closed topological 4-manifolds). These equations happen to be
written in universal covering spaces of the above manifolds. It is
shown that "long derivatives" in Dirac equation for a hydrogen atom
can be considered as covariant derivatives of spinors in the Weyl
noneuclidean 4-space and that electromagnetic potentials can be
considered as connectivities in this space. The gauge invariance of
electromagnetic field proves to be a natural consequence of the
basic principles of the proposed geometrical interpretation. Within
the suggested concept, atoms have no inside any point-like particles
(electrons) and this can give an opportunity to overcome the
difficulties of atomic physics connected with the many-body problem.

There were many attempts to geometrize electromagnetic field and
find out new interpretation of quantum mechanics formalism. The
distinctive feature of this work is that it combines geometrization
of electromagnetic field and geometrization of material field within
the unique topological idea. Some preliminary results were published
earlier [1-3].
\newpage
{\bf 1. Geometrization of the Dirac equation for a free
particle}

This equation has the form [4],
$$i\gamma_{\mu}\partial_{\mu}\psi=m\psi, \eqno(1)$$ where
$\partial_{\mu}=\partial/\partial x_{\mu},\quad \mu =1,2,3,4,\quad
\gamma_{\mu}$ are the Dirac matrices, $\psi (x)$ is the Dirac
bispinor, $x_{1}=t, x_{2}=x, x_{3}=y$, and $x_{4}=z$. There is
summation over repeating indices with a signature $(1,-1,-1,-1)$.
Here, $\hbar =c=1$. The solution to Eq.(1) has the form of a plane
wave
$$\psi=\psi_{p}\exp (-ip_{\mu}x_{\mu}),\eqno(2)$$ where the 4-momentum $p_{\mu}$
satisfies the relation
$$p_{1}^{2}-p_{2}^{2}-p_{3}^{2}-p_{4}^{2}=m^{2}.\eqno(3)$$

It is be helpful to rewrite Eq.(1) in the form
$$
(l_{1}^{-1}\gamma_{1}T_{1}-l_{2}^{-1}\gamma_{2}T_{2}-l_{3}^{-1}\gamma_{3}T_{3}-l_{4}^{-1}\gamma_{4}T_{4})\psi
=l_{m}^{-1}\psi,\eqno(4)$$ where operators $T_{\mu} (\mu =1,2,3,4)$
have the form
$$T_{\mu}=i(2\pi)^{-1}l_{\mu}\partial_{\mu},\quad
l_{\mu}=2\pi p_{\mu}^{-1},\quad l_{m}=2\pi m^{-1}.\eqno(5)$$ We
rewrite solution (2) as
$$\psi=\psi_{p}\exp (-2\pi
ix_{1}l_{1}^{-1}+2\pi ix_{2}l_{2}^{-1}+2\pi ix_{3}l_{3}^{-1}+2\pi
ix_{4}l_{4}^{-1}).\eqno(6)$$ We also rewrite the relation (3) as
$$l_{1}^{-2}-l_{2}^{-2}-l_{3}^{-2}-l_{4}^{-2}=l_{m}^{-2}.\eqno(7)$$
Note that all quantities in Eqs (4--7) has the dimensionality of
length.

Operators $T_{\mu}$ in Eq.(5) and functions $\psi (x)$ in Eq.(6) are
related by the following equation.
$$\psi^{'}(x^{'}_{\mu})\equiv T_{\mu}\psi(x_{\mu}^{'})=\psi(x_{\mu}),\quad
x_{\mu}^{'}=x_{\mu}+l_{\mu}.\eqno(8)$$ Within the group theory,
relation (8) means that $T_{\mu}$ is the operator representation of
a discrete group of parallel translations along $x_{\mu}$ axis with
the generator $l_{\mu}$. Relation (8) means also that the function
$\psi (x)$ is a basic vector of this representation [5].

In addition, being a four-component spinor, $\psi (x)$ is related to
the four-row Dirac matrices $\gamma_{\mu}$ by the equations [6]
$$\psi^{'}(x^{'})=\gamma_{\mu}\psi (x),\eqno(9)$$ where $x\equiv
(x_1,x_2,x_3,x_4)$,  $x^{'}\equiv (x_1,-x_2,-x_3,-x_4)$ for $\mu
=1$, and $x^{'}\equiv (-x_1,x_2,-x_3,-x_4)$ for  $\mu =2$, and so
on. Within the group theory, Eqs.(9) mean that the matrices
$\gamma_{\mu}$ are a matrix representation of the group of
reflections along three axes perpendicular to the $x_{\mu}$ axis.
Relation (9) means also that the function $\psi (x)$ is a basic
vector of the above symmetry-group representation.

A parallel translation with simultaneous reflection in the direction
perpendicular to the translation is often spoken of as "sliding
symmetry" [7]. Thus, we see that the operators
$$P_{\mu}=\gamma_{\mu}T_{\mu} \eqno(10)$$
form the representations of sliding symmetry group (sliding
symmetries in the $0x_{\mu}$ directions). Using the above notation,
we can rewrite Eq.(1) as
$$l_{\mu}^{-1}P_{\mu}\psi=l_{m}^{-1}\psi,\eqno(11)$$ where, as we
established before, solution (2) to this equation is a basic vector
of the above sliding symmetry group representation.

The Dirac bispinor (being or not being a solution of some physical
equation) is a special tensor that realizes one of the possible
representations of the Lorentz group (qroup of 4-rotations in the
Minkovskii space-time). We see, however, that the Dirac bispinor
$\psi (x)$, {\it being the solution (2) of Eq.(1)}, realizes also a
representation of some additional symmetry group in the Minkovskii
space---the sliding symmetry group. The physical Minkovskii
space-time has no sliding symmetry. So, we suggest that Eq.(1) is,
in fact, written not in the physical 4-space but in the auxiliary
pseudoeuclidean 4-space with the discrete sliding
symmetry---universal covering space of a closed topological
space-time 4-manifold. Such auxiliary spaces are used in topology
for the mathematical description of manifolds because the discrete
group operating in universal covering spaces coincide with the
so-called fundamental group of closed manifolds. The fundamental
group is a group of different classes of closed paths starting and
ending at the same point of a manifold, and this group is one of the
important topological characteristics of manifolds [8].

Thus, our suggestion is that the "discovered" sliding symmetries
along the four axes  are symmetries of a four-dimensional universal
covering space of a closed space--time manifold. It means that the
fundamental group of this manifold is a group of four sliding
symmetries. These four sliding symmetries correspond to four
possible classes of closed paths. The Dirac equation (1) is supposed
to be written in the above manifold covering space, and Eq.(11) has
to be considered as the group-theoretical relation that leads to the
restriction (7) for the four possible values of $l_{\mu}$ ($l_{\mu}$
is a length of the oriented closed path for one of the four possible
classes). Above manifold looks formally like a two-dimensional Klein
bootle---nonorientable manifold whose fundamental group is generated
by two sliding symmetries on the euclidean universal covering plane
[7].)

Thus, our hypothesis is that the Dirac Eq.(1) describes a space-time
closed 4-manifold and it would appear reasonable to identify this
manifold with a quantum particle describing by the same equation.
Being spatially bounded in the euclidean 3-space, the above closed
4-manifold extends along $t$-axis from $t=-\infty$ to $t=+\infty$ in
the pseudoeuclidean Minkovskii space (for example, a closed circle
in the two-dimensional pseudueuclidean space looks like an
equilateral hyperbola on the euclidean plane). Closed topological
manifold has no definite geometrical shape and its characteristic
size in our case is of the order  $l_{m}= \hbar/mc \sim 10^{-11} cm$
if $m$ is the electron mass. This particle could be imagined as a
some kind of neutrinos (majoran's or dirac's neutrinos) or as a some
kind of unknown neutral fermions that have only gravitational charge
$m$.

If the topological concept suggested in this work is true, then it
will lead to total modification of the quantum mechanics traditional
interpretation. Within the suggested geometrical approach, there is
no place for any classical mechanics notions: for example,
4-momentum components $p_{\mu}$ are replaced, according to (5), by
the inverse of length parameters $2\pi l^{-1}$ and so on. On the
other hand, there are no hidden parameters in the above concept (the
presence of such parameters is usually considered as the objection
against any possible new interpretation of the quantum mechanics
formalism [9,10]). At this stage, the suggested interpretation of
the Dirac equation (1) is no more than "kinematic" hypothesis. To be
the truth, it should be verified within dynamic problems and in the
next section we start with the simplest dynamic problem when the
interaction is the interaction with a given static field.

{\bf 2. Geometrization of a hydrogen atom}

Let us consider an electron interacting with a given electromagnetic
field in hydrogen atom. We neglect the hyperfine structure effects,
after which the Dirac equation takes the form [4]
$$i\gamma_{\mu}(\partial_{\mu}-ieA_{\mu})\psi=m\psi.\eqno (12)$$
Here $e$ и $m$ are the charge and mass of an electron, respectively
$A_{2}=A_{3}=A_{4}=0, A_{1}=-e/r$. Let us show that Eq.(12) can be
also considered as a relation written in some noneuclidean universal
covering space of a closed topological 4-manifold.

First of all, it was shown by Weyl that the expression in (12)
$$(\partial_{\mu}-ieA_{\mu}) \eqno (13)$$
can be considered as a covariant derivative of a tensor field in the
special noneuclidean space (Weyl space) and that $ieA_{\mu}$ can be
considered as a connectivity of this space [11]. This result was
generalized for Dirac bispinors $\psi (x)$ by Fock, who showed that
the expression
$$(\partial_{\mu}-ieA_{\mu})\psi \eqno (13)$$
can be considered as a covariant derivative of $\psi (x)$ in a
special case of the Weyl space ("planar" Weyl space) [12].
Therefore, we can consider Eq.(12) as a differential relationship
written in the Weyl space, where $ieA_{\mu}$ play the role of this
space connectivities.

Since the above results play a key role let us discuss the planar
Weyl space properties in more detail. This space geometry is
specified by linear and quadratic forms [9]
$$ds^{2}=g_{ik}dx_{i}dx_{k}=\lambda (x)
(dx^{2}_{1}-dx^{2}_{2}-dx^{2}_{3}-dx^{2}_{4}),\eqno (14)$$
$$d\varphi=\varphi_{\mu}dx_{\mu},\eqno (15),$$
where $\lambda (x)$--is an arbitrary differentiable positive
function of coordinates $x_{\mu}$. This space is invariant with
respect to the scale (or gauge) transformations
$$g_{ik}^{'}=\lambda g_{ik},\quad \varphi_{i}^{'}=\varphi_{i}-\frac
{\partial \ln \lambda}{\partial x_{i}}.\eqno (16)$$ Therefore, a
single-valued, invariant sense has not $\varphi_{i}$ but the
quantity (scale curvature)
$$F_{ik}=\frac {\partial \varphi_{i}}{\partial x_{k}}-\frac {\partial
\varphi_{k}}{\partial x_{i}}.\eqno (17)$$ The antisymmetric
propriety of the tensor $F_{ik}$ leads to the equations that are
analogous to the first pair of Maxwell's equations
$$\partial_{i}F_{kl}+\partial_{k}F_{li}+\partial_{l}F_{ik}=0.$$

This analogy and the gauge invariance of  $\varphi_{i}$ (like the
gauge invariance of electromagnetic potentials) lead Weyl to the
natural idea that vectors $\varphi_{i}$ can be identified with the
electromagnetic potentials and that tensor $F_{ik}$ can be
identified with the tensor of electromagnetic field strengths
$$\varphi_{\mu}\equiv ieA_{\mu}, \quad A_{\mu}^{'}=
A_{\mu}-\partial_{\mu}\chi, \quad \chi = ie\ln \lambda.\eqno (18)$$
Then (like in general relativity), Weyl attempted to identify the
geometry of his space (curvature and so on) with the geometry of a
real space-time distorted by the presence of electromagnetic field
[9]. But this hypothesis was contradictory to some observable
proprieties of the real physical space-time (it was shown by
Einstein in the supplement to the Weyl publication [9]), and the
Weyl's results were afterwards considered as methodical ones. In
contrast to Weyl, we propose that covariant derivative (13) is
written not in the real space-time but in the auxiliary space---
universal covering space of a topological manifold. So, there are no
objections against the Weyl space within our consideration. It means
that we can suggest that expression (13) is a covariant derivative
written in the Weyl, space and the 4-potentials $ieA_{\mu}$ play the
role of connectivities in this space.

Now (by analogy with the consideration in Section 1), we suppose
that the above-mentioned Weyl space is a universal covering space of
some 4-manifold that represents a hydrogen atom. The distinctive
feature of any universal covering space is the presence of a
discrete group operating in this space (isomorphic with the manifold
fundamental group)[8]. As for the equation for a free particle, this
propriety follows from the symmetry proprieties of solutions to
Eq.(12). For the atomic stationary state, the wave function has an
exponential factor
$$\exp (-iEt)=\exp (-l_{1}^{-1}x_{1}),\quad x_{1}=ct, \quad l_{1}=\hbar c/E.
$$ It means that the sliding symmetry (considered in Section 1)
occurs now only along the $x_{1}$ axis. The translational symmetries
along space axes are replaced by the symmetries of the rotation
group.

Thus, Eq.(12) can be interpreted as relation describing symmetry
properties of a closed topological 4-manifold. This differential
relation is written in the universal covering space of this
manifold, and this space is a noneuclidean Weyl space. The Coulomb
potential $A_{1}(x_{\mu})$ plays the role of this space connectivity
and the tensor of electric and magnetic fields $F_{ik}(x_{\mu})$
plays the role of this space curvature tensor.

In closing of this Section, we notice that the gauge invariance of
electromagnetic field in the hydrogen atom is a direct consequence
of a gauge invariance of the Weil space that is considered here as
the manifold universal covering space. Thus, the gauge invariance
idea treats in this work not as some additional principle but as a
natural consequence of the geometrical approach.

{\bf 3. Conclusion and final remarks}

It is shown that two basic quantum mechanical equations can be
interpreted as group-theoretical descriptions of specific
microscopic distortions of the space geometry from the euclidean
one. The gauge invariance of electromagnetic field in the hydrogen
atom proves not to be some additional theoretical principle,  but it
is a natural consequence of the basic principles of the proposed
geometrical interpretation. The real significance of the proposed
geometrical approach will be, of course, clear after its application
for handling unsolved physical problems. We are now working in this
direction and we look forward to the colleagues assistance.

In conclusion, notice one of the possible opportunities. Within the
suggested concept, atoms are considered as geometrical objects
without any point like particles (electrons), and it is hoped that
such objects can be described by functions of one space-time
variable. It, may be, means that the geometrical description of
many-electron atoms will give the chance to overcome difficulties of
a many-body problem.

\par\medskip

\noindent 1. O.A. Olkhov, in Proceedings of Int.Workshop on High
Energy Physics and Field Theory, Protvino, Russia, June 2001, p.327;
in Proceedings of "Photon 2001", Ascona, Switzerland, September
2001, p.360; in Proceedings of Int.Conf. on New Trends in
High-Energy Physics, Yalta (Crimea), September 2001, p.324; in Proceedings of "Group 24", Paris, France, July 2002. p. 363.\\
2. O.A. Olkhov, Moscow Institute of Physics and
Technology,Preprint  No.2002-1, Moscow,2002.\\
3. O.A. Ol'khov, hep-th/0511118\\
4. J.D. Bjorken and S.D. Drell, \emph{Relativistic Quantum Mechanics}. (McGraw-Hill Book Company, 1964).\\
5. M. Hamermesh, {\it Group Theory and Its Application to Physical
Problems}. (Argonne National Laboratory, 1964).\\
6. A.I. Achiezer, S.V. Peletminski, \emph{Fields and Fundamental
Ineractions}. (Naukova Dumka, Kiev, 1986), Ch.1\\
7. H.S.M. Coxeter, {\it Introduction to Geometry}. (John Wiley and Sons,Inc., N.Y.-London, 1961).\\
8. А.S. Schvartz, {\it Quantum Field Theory and Topology}. (Nauka,
Moscow, 1989.\\
9. J.S. Bell, Rev.Mod.Phys. \textbf{38}, 447 (1966)\\
10. J.V. Neumann, \emph{Mathematische grundlagen der
quantenmechanik}. (Springer, Berlin, 1935).\\
11. H.  Weyl, {\it Gravitation und Electrisitat}, Preus.Akad.Wiss,
Berlin (1918).\\
12. V. Fock, Zs. f. Phys., \textbf{57}, 261 (1929).\\

\noindent *E-mail address: olega@gagarinclub.ru\\

\end{document}